\documentclass[11pt,epsf]{article}
\pdfoutput=1
\RequirePackage{snapshot} 
 \usepackage{amsmath}
 \usepackage{graphicx}
 \usepackage[merge,numbers,compress]{natbib}
 \usepackage[T1]{fontenc}
 \usepackage{booktabs}
 \usepackage{xcolor} 
 \usepackage{xspace}
 \usepackage{dcolumn}
 \usepackage{hyperref}
 \usepackage{caption}
 \usepackage
 [subrefformat=parens,position=top,skip=-15pt,margin=15pt,justification=justified,singlelinecheck=false]
 {subcaption}

\newcommand{\fb}{{\ensuremath\unskip\,\text{fb}}\xspace}

\def\be{\begin{equation}}
\def\ee{\end{equation}}

\newcommand{\PH}{\ensuremath{\text{H}}\xspace}
\newcommand{\Pj}{\ensuremath{\text{j}}\xspace}
\newcommand{\Pp}{\ensuremath{\text{p}}\xspace}

\newcommand{\Pt}{\ensuremath{\text{t}}\xspace}

\newcommand{\PW}{\ensuremath{\text{W}}\xspace}
\newcommand{\PZ}{\ensuremath{\text{Z}}\xspace}

\newcommand{\Mt}{\ensuremath{m_\Pt}\xspace}

\newcommand{\MWOS}{\ensuremath{M_\PW^\text{OS}}\xspace}
\newcommand{\MW}{\ensuremath{M_\PW}\xspace}
\newcommand{\MZOS}{\ensuremath{M_\PZ^\text{OS}}\xspace}
\newcommand{\MZ}{\ensuremath{M_\PZ}\xspace}

\newcommand{\GH}{\ensuremath{\Gamma_\PH}\xspace}

\newcommand{\GZOS}{\ensuremath{\Gamma_\PZ^\text{OS}}\xspace}

\newcommand{\GWOS}{\ensuremath{\Gamma_\PW^\text{OS}}\xspace}

\newcommand{\GeV}{\ensuremath{\,\text{GeV}}\xspace}
\newcommand{\TeV}{\ensuremath{\,\text{TeV}}\xspace}

\newcommand{\alphas}{\ensuremath{\alpha_\text{s}}\xspace}

\newcommand{\GF}{\ensuremath{G_\mu}}

\newcommand{\ptsub}[1]{\ensuremath{p_{\text{T},#1}}\xspace}

\newcommand{\MVOS}{\ensuremath{M_{V}^\text{OS}}\xspace}%
\newcommand{\GVOS}{\ensuremath{\Gamma_{V}^\text{OS}}\xspace}%

\newcommand{\newc}{\newcommand}
\newc{\bi}{\begin{itemize}}
\newc{\ei}{\end{itemize}}
\newc{\benu}{\begin{enumerate}}
\newc{\eenu}{\end{enumerate}}
\newc{\bc}{\begin{center}}
\newc{\ec}{\end{center}}
\newc{\bfig}{\begin{figure}}
\newc{\efig}{\end{figure}}
\newc{\qbar}{\bar{q}}
\newc{\go}{\tilde{g}}
\newc{\PB}{\textsc{Powheg-Box}}

\newcommand{\Recola}{{\sc Recola}\xspace}

\newcommand{\proVBFHH}{{\sc proVBFHH}\xspace}

\newcommand{\MoCaNLO}{{\sc MoCaNLO}\xspace}

\newcommand{\collier}{{\sc Collier}\xspace}

\newcolumntype{.}{D{.}{.}{-1}}
\newcolumntype{d}[1]{D{.}{.}{#1}}

\colorlet{tableoverheadcolor}{gray!37.5}
\colorlet{tableheadcolor}{gray!25}
\colorlet{tablerowcolor}{gray!12.5}

\newlength{\width}
\newlength{\height}

\def\draftdate{\relax}
\def\mda{\relax}
\def\mua{\relax}
\def\mla{\relax}
\def\draft{
\def\thtystars{******************************}
\def\sixtystars{\thtystars\thtystars}
\typeout{}
\typeout{\sixtystars**}
\typeout{* Draft mode!
         For final version remove \protect\draft\space in source file *}
\typeout{\sixtystars**}
\typeout{}
\def\draftdate{\today}
\def\mua{\marginpar[\boldmath\hfil$\uparrow$]%
                   {\boldmath$\uparrow$\hfil}\color{black}%
                    \typeout{marginpar: $\uparrow$}\ignorespaces}
\def\mda{\color{red}\marginpar[\boldmath\hfil$\downarrow$]%
                   {\boldmath$\downarrow$\hfil}%
                    \typeout{marginpar: $\downarrow$}\ignorespaces}
\def\mla{\marginpar[\boldmath\hfil$\rightarrow$]%
                   {\boldmath$\leftarrow $\hfil}%
                    \typeout{marginpar: $\leftrightarrow$}\ignorespaces}
\def\Mua{\marginpar[\boldmath\hfil$\Uparrow$]%
                   {\boldmath$\Uparrow$\hfil}\color{black}%
                    \typeout{marginpar: $\uparrow$}\ignorespaces}
\def\Mda{\color{red}\marginpar[\boldmath\hfil$\Downarrow$]%
                   {\boldmath$\Downarrow$\hfil}%
                    \typeout{marginpar: $\downarrow$}\ignorespaces}
\def\Mla{\marginpar[\boldmath\hfil\textcolor{red}{$\Rightarrow$}]%
                   {\boldmath\textcolor{red}{$\Leftarrow $}\hfil}%
                    \typeout{marginpar: $\leftrightarrow$}\ignorespaces}
\overfullrule 5pt
\oddsidemargin 15mm
\marginparwidth 29mm
}

\begin{document}

\title{\hfill ~\\[-30mm]
\phantom{h} \hfill\mbox{\small Cavendish-HEP 20/04, OUTP-20-03P, ZU-TH 17/20}
\\[1cm]
\vspace{13mm}   \textbf{Precise predictions for double-Higgs production \\ via vector-boson fusion}}

\date{}
\author{
Fr\'ed\'eric A. Dreyer$^{1\,}$\footnote{E-mail:  \texttt{frederic.dreyer@physics.ox.ac.uk}} ,
Alexander Karlberg$^{1\,}$\footnote{E-mail:  \texttt{alexander.karlberg@physics.ox.ac.uk}} ,
Jean-Nicolas Lang$^{2\,}$\footnote{E-mail:  \texttt{jlang@physik.uzh.ch}} ,
Mathieu Pellen$^{3\,}$\footnote{E-mail:  \texttt{mpellen@hep.phy.cam.ac.uk}}
\\[9mm]
{\small\it $^1$Rudolf Peierls Centre for Theoretical Physics, University of Oxford,}\\
{\small\it Clarendon Laboratory, Parks Road, Oxford OX1 3PU, United Kingdom} \\[3mm]
{\small\it $^2$Universit\"at Z\"urich, Physik-Institut,}\\
{\small\it CH-8057 Z\"urich, Switzerland} \\[3mm]
{\small\it $^3$ University of Cambridge, Cavendish Laboratory,} \\ %
{\small\it 19 JJ Thomson Avenue, Cambridge CB3 0HE, United Kingdom}\\[3mm]
        }
\maketitle

\begin{abstract}
\noindent

Theoretical predictions with next-to-next-to-leading order (NNLO) QCD
accuracy combined with the next-to-leading order (NLO) electroweak
(EW) corrections are presented for differential observables of the
double-Higgs production process via vector-boson fusion.
While the QCD corrections were previously known, the EW ones are
computed here for the first time.
The numerical results are obtained for a realistic experimental set-up
at the LHC and are presented in the form of fiducial cross sections
and differential distributions.
Within this setup we find that the VBF approximation employed in the
NNLO QCD correction is accurate at the sub-percent level.
We find that the NLO EW corrections within the fiducial volume are
$-4.8\%$, making them of almost the same order as the NLO QCD
corrections. In some kinematic regions they can grow as large as
$-30\%$ making them the dominant radiative corrections.
When the EW corrections are combined with the NNLO QCD corrections we
find a total correction of $-14.8\%$.
The results presented here thus comprise the state-of-the-art
theoretical predicition for the double-Higgs production via
vector-boson fusion, which will be of value to the high-luminosity
programme at the LHC.

\end{abstract}
\thispagestyle{empty}
\vfill

\newpage

\section{Introduction}

The discovery of the Higgs boson at the Large Hadron Collider
(LHC)~\cite{Aad:2012tfa,Chatrchyan:2012xdj} opened the door to a new
era in high-energy physics, dominated by the quest for precision, and
which will culminate in the next decade with the ambitious
High-Luminosity programme at the LHC~\cite{Azzi:2019yne}.
The determination of all the fundamental parameters of the Higgs
sector will be at the heart of this programme.
Beyond the mass and the width of the Higgs boson, its interactions
with the other particles of the Standard Model (SM) will be precisely
scrutinised.
One of the most far-reaching tasks will be the determination of the
Higgs self-couplings.
These are fundamental parameters of the SM Lagrangian that determine
the shape of the Higgs potential.
The investigation of these properties will rely notably on the
investigation of processes involving a pair of Higgs bosons in the
final
state~\cite{Aaboud:2018sfw,Aaboud:2018ewm,Aaboud:2018ftw,Aaboud:2018knk,Aaboud:2016xco,Aad:2015xja,Aad:2015uka,Aad:2014yja,Sirunyan:2018tki,Sirunyan:2018iwt,Sirunyan:2017guj,Sirunyan:2017djm,Sirunyan:2017tqo}.

The production of two Higgs bosons via vector-boson fusion (VBF)
\emph{i.e.}\ $\Pp\Pp \to \PH\PH\Pj\Pj$ is a particularly important
process for the determination of the triple-Higgs
coupling~\cite{Baglio:2012np}.
While gluon fusion is the dominating double-Higgs production
mode~\cite{deFlorian:2016spz,deFlorian:2013jea,deFlorian:2016uhr,deFlorian:2015moa,Grazzini:2018bsd},
the VBF channel offers unique opportunities for measurements of
Higgs pair production.
In VBF, the Higgs bosons are produced at leading order from heavy
gauge bosons that are themselves radiated off two quark lines.
These two quarks offer a useful handle as tagging jets for its
experimental measurement, providing a promising channel for studies of
the trilinear and quartic Higgs couplings at the
LHC~\cite{Baglio:2012np,Bishara:2016kjn}.

While single-Higgs production via VBF has been measured~\cite{Khachatryan:2016vau},
its double-Higgs counterpart is not yet accessible with the current
experimental data set.
The reason is that the cross section at $14\TeV$ is of the order of $1 \fb$ as it
is a purely electroweak (EW) process and requires very exclusive event
selections in order to single it out from its background.
Nonetheless, this process is already used to look for new
physics~\cite{Aad:2020kub}.
The High-Luminosity programme of the LHC is aiming at collecting about
$3000\fb^{-1}$ in the next two decades.
This should allow the observation of the double-Higgs production via VBF.
To that end, precise and reliable theoretical predictions are required.

In this article we provide for the first time next-to-leading order
(NLO) EW corrections and combine them with the existing
next-to-next-to-leading order (NNLO) QCD corrections.
Both types of corrections are very different in magnitudes and shapes
as they account for different physical effects.
The QCD ones are usually larger and their missing higher-order terms
can be well estimated by a variation of the renormalisation and
factorisation scales.
In Refs.~\cite{Dreyer:2018rfu,Dreyer:2020urf}, it has been shown that
only NNLO QCD corrections provide reliable predictions at the
differential level.
Recently, the ${\rm N}^3{\rm LO}$ corrections to the inclusive cross
section have also been computed, and were shown to be at the few
permille level~\cite{Dreyer:2018qbw}.
The EW corrections on the other hand are typically rather suppressed
and appear in the high-energy tail of differential distributions where
the effect of Sudakov logarithms become large.

The present article is organised as follow: in the first part, the
details of the computation are provided.  In particular the numerical
inputs and the event selection are explained.
In the second part, the results are given in the form of cross
sections and differential distributions.
Finally, the conclusion contains a short summary of the article as
well as concluding remarks.

\section{Details of the computation}

\subsubsection*{Description of the process}

\begin{figure}
  \centering
        \setlength{\parskip}{2pt}
        \begin{subfigure}{0.33\textwidth}
                \subcaption{}
                \vspace{6mm}
                  \includegraphics[width=0.95\textwidth]{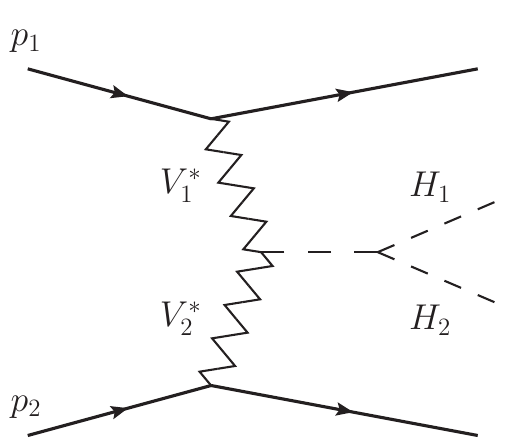}
                \label{fig:diagLO}
        \end{subfigure}%
        \hfill
        \begin{subfigure}{0.33\textwidth}
                \subcaption{}
                \vspace{6mm}
                 \includegraphics[width=0.95\textwidth]{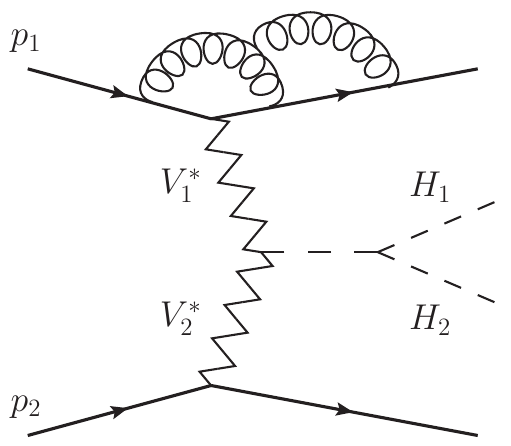}
                \label{fig:diagQCD}
        \end{subfigure}%
        \hfill
        \begin{subfigure}{0.33\textwidth}
                \subcaption{}
                \vspace{6mm}
                 \includegraphics[width=0.95\textwidth]{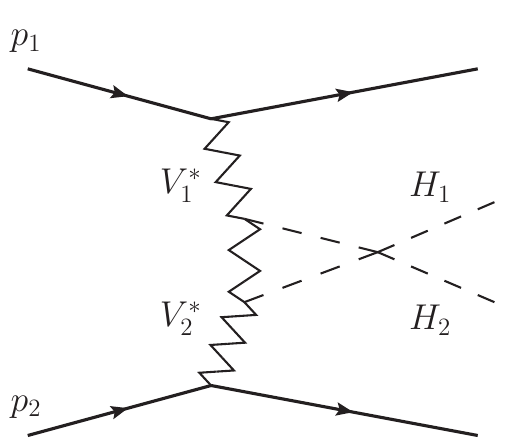}
                \label{fig:diagEW}
        \end{subfigure}
        \caption{Examples of feynman diagrams contributing to the VBF
          Higgs pair production process at LO (a), NNLO QCD
          (b) and NLO EW (c).}
  \label{fig:diags}
\end{figure}

The process of interest is the double-Higgs production via VBF, which
can be expressed as
\begin{equation}
 \Pp\Pp \to \PH \PH \Pj \Pj .
\end{equation}
It is a purely EW process which is defined at tree level at the order
$\mathcal{O}\left(\alpha^4 \right)$.
At this order, it contains the VBF topology which is defined as two
quark lines that radiate heavy gauge bosons that fuse to give rise to
two Higgs bosons, as shown in Fig.~\ref{fig:diagLO}.
Alternatively, the two Higgs bosons can be radiated from a heavy
gauge boson decaying subsequently into two quarks.
The latter topology is usually referred to as Higgs Strahlung.
While these two topologies are intimately related, they entail rather
different physical effects.

It has lead, on the experimental side, to consider the two processes
separately by imposing rather stringent cuts on the invariant mass and
the rapidity difference of the two tagging jets.
This has motivated the use of the VBF approximation for theoretical
predictions, which neglects s-channel contributions as well as t/u
interferences in a gauge-invariant way.
Additionally the two quark lines are not allowed to exchange virtual
or real gluons.
One advantage of this approach is that it greatly simplifies the
computations.
In particular, it has allowed the computation of QCD corrections up to
N$^3$LO of single and double Higgs
production~\cite{Figy:2008zd,Bolzoni:2010xr,Liu-Sheng:2014gxa,Cacciari:2015jma,Dreyer:2016oyx,Cruz-Martinez:2018rod,Dreyer:2018rfu,Dreyer:2018qbw}.
The differential cross section of the di-Higgs process was also
computed at NLO with matching to parton showers in this
approximation~\cite{Frederix:2014hta}.

In the present article, we have used the full computations at LO, NLO
QCD, and NLO EW while we have utilised the VBF approximation at NNLO.
In order to justify their combination, we have used a rather exclusive
experimental set-up where the VBF approximation holds at the per-mille
level.
In addition, we have added a correction factor to the NNLO corrections
to account for the differences between the full and the VBF
computation.
More precisely we have checked that the difference between the full
computation and the VBF-approximate one is identical at LO and NLO QCD in all
differential distributions.
Recently, extensive studies investigating the quality of such
approximation at NLO in QCD for vector-boson scattering
(VBS)~\cite{Ballestrero:2018anz} and the EW production of a Higgs
boson in association with three jets~\cite{Campanario:2018ppz} have
been performed.
We have refrained from presenting similar results for the process at hand
as it is very close to processes mentioned above.
Instead we have displayed the correction factor used in all differential distributions which indicates the difference between the full and the approximate computation.

\begin{itemize}
\item QCD corrections \\
  In the full computation at NLO, the real corrections consist in all
  the contributions of the type $\Pp\Pp \to \PH \PH \Pj \Pj \Pj$ at
  order $\mathcal{O}\left(\alphas \alpha^4 \right)$.
  The virtual corrections entail all the possible gluon insertions on
  a given quark line, with gluon exchanges between the two quark lines
  vanishing for colour reasons.
  At NNLO, for which an example diagram is shown in
  Fig.~\ref{fig:diagQCD}, heavy-quark loop-induced diagrams are
  neglected, as well as $t$-/$u$-channel interferences, single-quark
  line contributions and loop induced interferences between VBF and
  gluon-fusion production.
  These contributions have been shown to be suppressed to less than a
  percent in the single-Higgs
  case~\cite{Ciccolini:2007ec,Harlander:2008xn,Bolzoni:2011cu}, but
  are not known for Higgs-pair production. Although they could be
  sizeable there is no apriori reason to expect them to be enhanced in
  di-Higgs production. A dedicated future study is however necessary
  to confirm whether or not these effects can be neglected. The
  non-factorisable diagrams involving the exchange of two gluons
  between the quark lines, have recently been estimated in
  Ref.~\cite{Dreyer:2020urf}.
  In this work we therefore also
  provide an estimate of the non-factorisable corrections, but will
  show them separately from the factorisable corrections. Unless
  explicitly specified, when referring to NNLO QCD corrections, we
  will always mean the factorisable ones. We compute the factorisable
  NNLO QCD corrections using the projection-to-Born method as detailed
  in Ref.~\cite{Dreyer:2018rfu}.
 
\item EW corrections \\
  For the EW corrections the real radiations are made
  of the $\Pp\Pp \to \Pj \Pj \PH \PH \gamma$ channels at order
  $\mathcal{O}\left(\alpha^5 \right)$.
  At the same order, the virtual corrections are obtained by inserting
  EW particles anywhere possible in the tree-level topologies, an
  example of which is shown in Fig.~\ref{fig:diagEW}.
  Note that at the order $\mathcal{O}\left(\alpha^5 \right)$,
  photon-induced contributions also arise.
  These have been neglected in the present work as these have been
  shown to be rather small for similar processes
  \cite{Biedermann:2017bss,Denner:2019tmn,Ciccolini:2007ec}.\footnote{Using the same phase-space cuts, 
  we have recomputed the photon-induced corrections with {\sc HAWK} \cite{Denner:2014cla} for single Higgs production via VBF using the NNPDF31\_nnlo\_as\_0118\_luxqed set \cite{Bertone:2017bme}.
  We have found that this contribution amounts to $0.9\%$ of the fiducial cross section.}
  Note that EW corrections to single-Higgs production have been computed for the first time in Refs.~\cite{Ciccolini:2007jr,Ciccolini:2007ec} and are available in {\sc HAWK} \cite{Denner:2014cla}.
  Later they have also been obtained in {\sc VBFNLO} \cite{Figy:2010ct,Baglio:2014uba}.
 
\end{itemize}

As mentioned previously, all LO and NLO predictions are based on the
full computation, \emph{i.e.}\ without employing the VBF approximation.
These have been obtained from the Monte Carlo \MoCaNLO, which has
already been used for a variety of processes and in particular
VBS ones \cite{Biedermann:2016yds,Biedermann:2017bss,Denner:2019tmn} at NLO EW
and NLO QCD.
The matrix elements are provided by \Recola
\cite{Actis:2016mpe,Actis:2012qn,Denner:2017wsf} which internally uses
the \collier library \cite{Denner:2014gla,Denner:2016kdg} to evaluate
tensor integrals.

On the other hand, the NNLO QCD corrections have been obtained from
\proVBFHH v1.2.0~\cite{Dreyer:2018rfu,Dreyer:2018qbw} which uses the
projection-to-Born method~\cite{Cacciari:2015jma} to compute the fully
differential NNLO corrections in the VBF approximation. In order to
correct for the mismatch between this computation and the full
computation used for the LO and NLO computations, we compute a differential
correction factor
\begin{align}
  K_{\mathrm{full/VBF}} = \frac{d\sigma^{\rm full}_{\rm LO}}{d\sigma^{\rm VBF}_{\rm LO}}
\end{align}

and obtain the NNLO cross section provided below in the following way

\begin{equation}
 \sigma_{\rm NNLO\; QCD} = \sigma^{\rm full}_{\rm LO} + \delta^{\rm full}_{\rm NLO\; QCD} + K_{\mathrm{full/VBF}}\delta^{\rm VBF}_{\rm NNLO\; QCD},
\end{equation}
where the \emph{full} quantities refer to the computations with no
approximations and the \emph{VBF} one to the relative NNLO corrections
in the VBF approximation.
At the differential level, the NNLO predictions are obtained in the
same way. We have checked numerically that if one were to obtain
$K_{\mathrm{full/VBF}}$ using instead the NLO cross section, the
results do not change within statistical uncertainties. This is only
true under the very stringent cuts that we will introduce below.

The non-factorisable NNLO QCD corrections,
$\delta^{\rm NF}_{\rm NNLO\; QCD}$, have also been obtained using
\proVBFHH v1.2.0 as computed in Ref.~\cite{Dreyer:2020urf}.
Since they are very different in nature from the factorisable
corrections, we do not correct them by $K_{\mathrm{full/VBF}}$, and
show them separately from all other results.

Finally, to combine the NNLO QCD prediction and the NLO EW ones, we
have followed the Higgs cross section working group recommendation
\cite{Heinemeyer:2013tqa} for VBF which reads

\begin{equation}
  \sigma_{\rm NNLO \; QCD \times NLO \; EW} = \sigma_{\rm NNLO \; QCD} \left(1 + \frac{\delta^{\rm full}_{\rm NLO \; EW}}{\sigma^{\rm full}_{\rm LO}} \right),
  \label{eq:NNLO-QCD-NLO-EW}
\end{equation}
with $\sigma^{\rm full}_{\rm NLO \; EW} = \sigma^{\rm full}_{\rm LO} + \delta^{\rm full}_{\rm NLO \; EW}$.
The same formula has been applied differentially.
With such a prescription we provide thus best predictions at fixed
order in the SM for double-Higgs production in VBF. We
stress that we do not include $\delta^{\rm NF}_{\rm NNLO\; QCD}$ in
this quantity, but rather quote it separately.

\subsubsection*{Numerical inputs}

Our predictions are obtained for proton-proton
collisions at the LHC running with a centre-of-mass energy of
$\sqrt{s} = 14 \TeV$.
The 5-flavour scheme is used throughout the computation \emph{i.e.}\
the bottom quarks are considered massless.  Bottom quarks are also
included in the jet definition.
For the parton distribution functions (PDF) and $\alpha_s$, the
NNPDF31\_nnlo\_as\_0118\_luxqed set \cite{Bertone:2017bme} has been
used for all computations \emph{i.e.}\ at LO, NLO, and NNLO as
implemented in LHAPDF6~\cite{Buckley:2014ana}.
The choice of the central renormalisation and factorisation scale is the same
as the one used in Ref.~\cite{Dreyer:2018rfu} and reads

\begin{equation}
 \mu = \sqrt{\frac{M_\PH}{2} \sqrt{\left(\frac{ M_\PH}{2} \right)^2 + \ptsub{\PH\PH}^2}} ,
\end{equation}
with $M_\PH$ the mass of the Higgs boson and $\ptsub{\PH\PH}$ the
transverse momentum of the di-Higgs system. 
In order to estimate the impact of missing higher-order QCD
corrections, we perform a 3-point scale variation defined by
$\mu_{\rm ren} = \mu_{\rm fac} = \{1/2,1,2\}\times\mu$.
Note that this method is applied for the LO predictions and the
factorisable QCD corrections only.
For the EW corrections, such a method does not provide a good estimate
of missing higher-order and thus NLO EW corrections are simply applied
as an overall factor.

The masses and widths used for the numerical simulations read
\begin{alignat}{2}
                  \Mt   &=  173.21\GeV,       & \quad \quad \quad M_\PH &=  125.0\GeV,  \nonumber \\
                \MZOS &=  91.1876\GeV,      & \quad \quad \quad \GZOS &= 2.4952\GeV,  \nonumber \\
                \MWOS &=  80.385\GeV,       & \GWOS &= 2.085\GeV.
\end{alignat}
The mass of the bottom quark is taken to be zero in accordance with
the 5-flavour scheme.
The width of the top quark is also taken to be zero as no top quarks
are produced resonantly.
The value taken for the Higgs-boson mass is taken from the report of
the Higgs cross section working group \cite{Heinemeyer:2013tqa}.
Note that the width has been taken to zero as the Higgs bosons are on-shell
external states.
They can nonetheless appear as internal propagator in the splitting
$\PH^* \to \PH \PH$.
As the invariant mass of the off-shell Higgs boson tends to be far
from the on-shell mass and since the value of the width,
$\GH = 4.07 \times 10^{-3}\GeV$, is small, setting the width to zero
has no numerical impact%
\footnote{For the NNLO corrections the width has been used for the internal Higgs propagators.}.
The pole masses and widths used for the simulations are obtained from
the measured on-shell (OS) values \cite{Bardin:1988xt} for the W and
Z~bosons according to
\begin{equation}
        M_V = \frac{\MVOS}{\sqrt{1+(\GVOS/\MVOS)^2}}\,,\qquad
\Gamma_V = \frac{\GVOS}{\sqrt{1+(\GVOS/\MVOS)^2}},
\end{equation}
with $V=\PW, \PZ$.

The $G_\mu$ scheme \cite{Denner:2000bj} is used for all computations
and is translated to $\alpha$ via
\begin{equation}
  \alpha = \frac{\sqrt{2}}{\pi} G_\mu \MW^2 \left( 1 - \frac{\MW^2}{\MZ^2} \right)  \qquad \text{and}  \qquad   \GF    = 1.16637\times 10^{-5}\GeV.
\end{equation}

In all the LO and NLO computations, the intermediate W/Z-boson
resonances are treated in the complex-mass scheme~\cite{Denner:1999gp,Denner:2005fg,Denner:2006ic} to ensure gauge
independence of all amplitudes.

\subsubsection*{Event selection}
\label{sec:cuts}
The experimental cuts are rather generic and typically intend to
single out VBF EW contributions from their QCD background.  To that
end, high invariant-mass and large rapidity difference between the two
tagging jets are required. The tagging jets are defined by requiring
that each jet fulfils the following condition:

\begin{align}
 \ptsub{\Pj} >  25\GeV \qquad {\rm and} \qquad |y_{\Pj}| < 4.5 .
\end{align}

The jets are clustered using the anti-$k_t$
algorithm~\cite{Cacciari:2008gp} with $R=0.4$, using
\texttt{FastJet v3.3.0}~\cite{Cacciari:2011ma} in the case of
\proVBFHH. The two hardest jets in the transverse momentum fulfilling
these requirements are required to obey the VBF-selection cuts which
read
\begin{align}
 m_{\Pj_1 \Pj_2} >  600\GeV \qquad {\rm and} \qquad |y_{\Pj_1} - y_{\Pj_2}| > 4.5.
\end{align}

Note that we do not apply a cut ensuring that both tagging jets are in
opposite hemispheres as in Ref.~\cite{Dreyer:2018rfu}.  Also, the
event selection is completely inclusive in the final state Higgs
bosons and no cuts of any kind are applied to them.

\section{Results}

In this section we show numerical results for cross sections and
differential distributions.
Given that the QCD corrections have already been presented in
Ref.~\cite{Dreyer:2018rfu,Dreyer:2020urf}, the discussion focuses more
on their combination with the EW ones which are presented here for the
first time.
Nonetheless, some distributions were not shown in
Ref.~\cite{Dreyer:2018rfu} and are therefore discussed here in more
details.
In the following, the ${\rm NNLO \; QCD \times NLO \; EW}$ predictions
are sometimes referred to as \emph{state-of-the-art predictions}.
Finally, the differences between the \emph{full} and the \emph{VBF}
computation are highlighted.

\begin{table*}[t] 
  \centering
  \phantom{x}\medskip
  \small{\begin{tabular}{cccc|cc}
    \toprule
    $\sigma^{\rm full}_{\rm LO}$ & $\delta^{\rm full}_{\rm NLO\; QCD}$ & $\delta^{\rm VBF}_{\rm NNLO\; QCD}$  & $\delta^{\rm full}_{\rm NLO \; EW}$ & $\sigma_{\rm NNLO \; QCD \times NLO \; EW}$ &$\delta^{\rm NF}_{\rm NNLO\; QCD}$ [fb] \\
    \midrule
    $0.78444(9)^{+0.0825}_{-0.0694}$ & $-0.07110(13)$ & $-0.0115(5)$ & $-0.0374(2)$ & $0.6684(5)^{+0.002}_{-0.0004}$ & $-0.001766(7)$ \\
    \midrule
    $\phantom{0.78444(9)}^{+10.5\%}_{-8.8\%}$ & $-9.1 \%$ & $-1.5 \%$ & $-4.8 \%$ & ${-14.8 \%}^{+0.3\%}_{-0.06\%}$ & $-0.23 \%$ \\
    \bottomrule
  \end{tabular}}
\caption{The fiducial cross section for the process
  $\Pp\Pp \to \PH \PH \Pj \Pj$, expressed in $\fb$ and in per cent,
  computed according to Eq.~\eqref{eq:NNLO-QCD-NLO-EW} at $14 \TeV$
  and under the selection cuts given in Sec.~\ref{sec:cuts}.
  The numbers in per cent are with respect to the LO cross section.
  The errors given in parenthesis are purely statistical whereas the
  additional uncertainties quoted for $\sigma^{\rm full}_{\rm LO}$ and
  $\sigma_{\rm NNLO \; QCD \times NLO \; EW}$ are the QCD scale
  variations.
  We also show $\delta^{\rm NF}_{\rm NNLO\; QCD}$ separately.
  The value of the correction factor to go from the VBF approximation
  to the full computation is $K_{\mathrm{full/VBF}}=0.99220(11)$.}
  \label{tab:xsec}
\end{table*}

In Table~\ref{tab:xsec}, fiducial cross sections and higher-order
corrections are displayed for the event selection presented in
Sec.~\ref{sec:cuts}.  They are expressed both in femto barn and in per
cent.  The numbers in per cent are with respect to the LO cross
section.  The numbers in parenthesis indicate the statistical error
while the additional information on $\sigma^{\rm full}_{\rm LO}$ and
$\sigma_{\rm NNLO \; QCD \times NLO \; EW}$ gives the scale variation
estimate.  Note that the total statistical uncertainty is not obtained
by adding the individual statistical uncertainties in quadrature, as
these are all correlated.

One of the main messages of Table~\ref{tab:xsec} is that the QCD
corrections are negative as for similar signatures such as single
Higgs-production via VBF or VBS at the LHC.  In
addition, the higher-order QCD corrections dramatically reduce the
uncertainty associated with missing QCD higher orders.  In particular,
it goes from $\left[+10.5\% , -8.8\%\right]$ at LO to
$\left[{+0.3\%}, {-0.06\%}\right]$ at NNLO in QCD.
We note that the non-factorisable NNLO QCD corrections are also negative but are about one order of magnitude smaller than the factorisables ones.

The second important point is the size of the EW corrections.
It has recently been found (and further confirmed in
Refs.~\cite{Denner:2019tmn,Chiesa:2019ulk}) that large EW corrections
are an intrinsic feature of VBS at the LHC
\cite{Biedermann:2016yds}.
It originates from the quantum numbers of the particles involved in
the process as well as the large scale induced by the massive
t-channel exchange \cite{Denner:1997kq}.
For such processes, the corrections reach about $-15\%$ to $-20\%$ of
the LO prediction.
On the other hand, for single-Higgs production via VBF, EW corrections
have been found to be around $-5\%$
\cite{Ciccolini:2007ec,Ciccolini:2007jr}.
It is thus interesting to observe that, despite having a higher
typical scale, the magnitude of the EW corrections for double-Higgs production via VBF is very close to
the single-Higgs one.
In particular, in VBS the typical scale (the invariant mass of the four leptons) is $\langle m_{4 \ell} \rangle\sim 390\GeV$ while the VBF case it is even larger with $\langle m_{\PH\PH} \rangle\sim 610\GeV$.
In the same way as in Ref.~\cite{Biedermann:2016yds}, one can derive a leading-logarithmic approximation using Ref.~\cite{Denner:2000jv}.
Because the quantum numbers of the Higgs boson, such as the effective EW Casimir operator (see Eq.~(B.10) in Ref.~\cite{Denner:2000jv}), are significantly smaller than the ones of the Z or W gauge bosons, the logarithm coefficients are reduced with respect to the VBS case.
For example, the coefficient of the double logarithms, which is directly proportional to the effective EW Casimir operator, is smaller by about a factor two.
This implies, that VBF does not feature intrinsic large EW corrections as VBS.

The QCD corrections on the other hand
tend to be somewhat smaller for double-Higgs production compared to
single Higgs. This is due to the larger energy transfer in the
t-channel which leads to harder jets and a higher dijet invariant
mass. This in turn means that fewer events are lost due to QCD radiation.
Overall, the state-of-the-art prediction displays a
correction of about $-15\%$ with respect to the LO prediction.
Finally, the numerical value of the correction factor is
$K_{{\rm full/VBF}} = 0.99220(11)$ at the level of the fiducial cross
section.
It means that for the (rather exclusive) fiducial volume chosen here,
the VBF approximation is reliable below the per-cent level.
As shown later, this correction factor is not constant over the
kinematic range and thus motivates its incorporation in our final
predictions.
    
\begin{figure}
        \setlength{\parskip}{-10pt}
        \begin{subfigure}{0.49\textwidth}
                \subcaption{}
                \vspace{3mm}
                 \includegraphics[width=\textwidth,page=1]{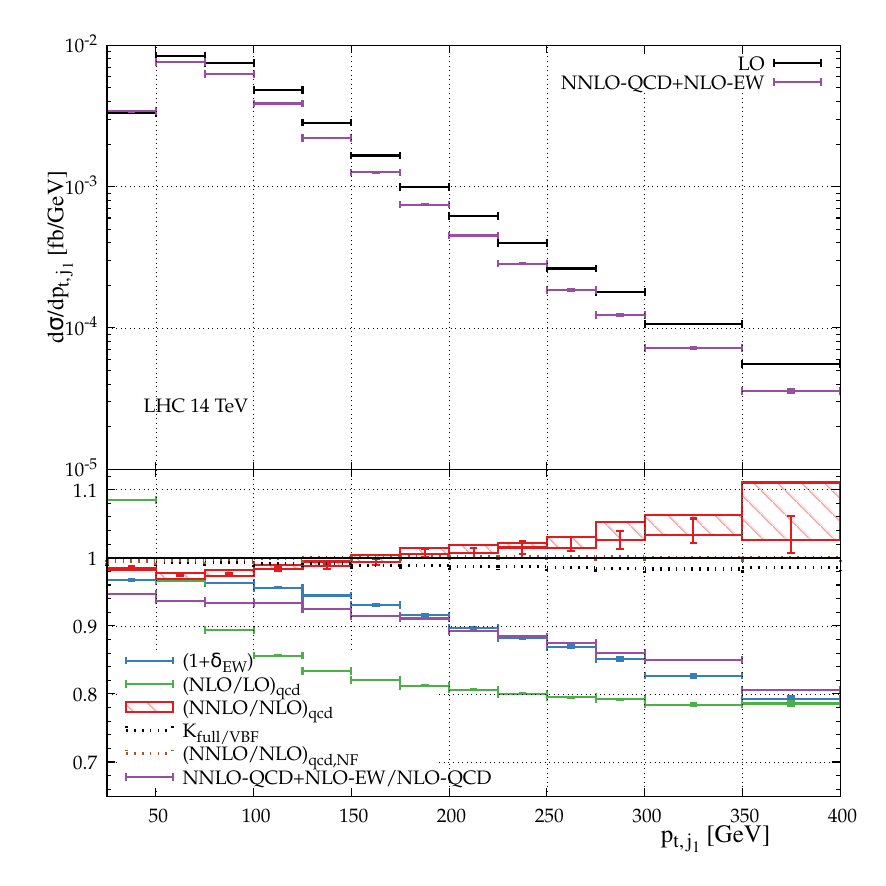}
                \label{plot:pTj1}
        \end{subfigure}
        \hfill
        \begin{subfigure}{0.49\textwidth}
                \subcaption{}
                \vspace{3mm}
                 \includegraphics[width=\textwidth,page=2]{results/VBFHH-plots-paper-extended.pdf}
                \label{plot:pTj2}
        \end{subfigure}

        \begin{subfigure}{0.49\textwidth}
                \subcaption{}
                \vspace{3mm}
                 \includegraphics[width=\textwidth,page=3]{results/VBFHH-plots-paper-extended.pdf}
                \label{plot:pTH1}
        \end{subfigure}
        \hfill
        \begin{subfigure}{0.49\textwidth}
                \subcaption{}
                \vspace{3mm}
                 \includegraphics[width=\textwidth,page=4]{results/VBFHH-plots-paper-extended.pdf}
                \label{plot:pTH2}
        \end{subfigure}

        \vspace*{-3ex}
        \caption{\label{fig:pT}%
          Differential distributions for $\Pp\Pp \to \Pj \Pj \PH \PH$
          at the LHC with centre-of-mass energy of $14\TeV$:
          \subref{plot:pTj1}~transverse momentum of the hardest
          jet~(top left), %
          \subref{plot:pTj2}~transverse momentum of the second hardest
          jet~(top right), \subref{plot:pTH1}~transverse momentum of
          the hardest Higgs boson~(bottom left), and
          \subref{plot:pTH2}~transverse momentum of the second hardest
          Higgs boson~(bottom right).  The upper panel shows the
          absolute contributions at NNLO QCD + NLO EW and the LO
          prediction.  The lower panel shows the relative
          corrections. The bands denote the envelope of the QCD scale
          variation. Note that the non-factorisable corrections to the
          transverse momenta of the jets should not be trusted at
          large values, as the underlying eikonal approximation breaks
          down.  }
\end{figure}

In Fig.~\ref{fig:pT}, several transverse-momentum distributions are
shown.
The first two are the transverse momentum of the hardest jet and
second hardest jet.
Comparing the EW corrections to the QCD ones, we observe that their
shapes are rather similar.
These corrections are driven by Sudakov logarithms that grow
negatively large towards the high-energy region.
In this case, high energy refers to the high transverse momenta.
For the hardest jet, it goes to about $-25\%$ at $400\GeV$, while for
the second hardest jet, the value $-25\%$ is reached already at
$200\GeV$.
Concerning the transverse momenta of the Higgs bosons, ordered by
their transverse momentum, the overall behaviour is similar and the
corrections grow smoothly towards higher momenta.
It is interesting to notice that in this case the difference between
the full and the VBF computation is of $10\%$ to $20\%$ at $400\GeV$.
This is the most striking effect of the VBF approximation that we have
observed in the present study.
But it happens in a rather suppressed part of the phase space.
For other distributions, the results are very much in line with
Ref.~\cite{Heinemeyer:2013tqa} where the difference between the full
and the VBF computation have been found to be small.

\begin{figure}
        \setlength{\parskip}{-10pt}
        \begin{subfigure}{0.49\textwidth}
                \subcaption{}
                \vspace{3mm}
                 \includegraphics[width=\textwidth,page=5]{results/VBFHH-plots-paper-extended.pdf}
                \label{plot:yj1}
        \end{subfigure}
        \hfill
        \begin{subfigure}{0.49\textwidth}
                \subcaption{}
                \vspace{3mm}
                 \includegraphics[width=\textwidth,page=7]{results/VBFHH-plots-paper-extended.pdf}
                \label{plot:yH1}
        \end{subfigure}
        \vspace*{-3ex}
        \caption{\label{fig:y}%
                Differential distributions for $\Pp\Pp \to \Pj \Pj \PH \PH$ at the LHC with centre-of-mass energy of $14\TeV$:
                \subref{plot:yj1}~rapidity of the hardest jet~(left) and
                \subref{plot:yH1}~rapidity of the hardest Higgs boson~(right).
                The upper panel shows the absolute contributions at NNLO QCD + NLO EW and the LO prediction.
                The lower panel shows the relative corrections. The bands denote the envelope of the QCD scale variation.}
\end{figure}

The rapidity distributions of the hardest jet and Higgs boson are shown in Fig.~\ref{fig:y}.
For the rapidity distribution of the hardest jet, the NLO QCD
corrections go from $-20\%$ at $0$ rapidity to about $+4\%$ at $4.5$
rapidity.
The NNLO QCD corrections are smaller and vary within $4\%$ across the
phase space.
The NLO EW corrections, on the other hand, hardly change over the
displayed rapidity range and essentially inherit the overall
normalisation.
The rapidity distribution of the hardest Higgs boson displays even more stable corrections.
The NLO QCD corrections fluctuate by less than $5\%$ over the whole
spectrum and the NNLO QCD corrections are also very stable (few per
cent variation).
The NLO EW corrections show a small shape distortion at the per-cent level.
In both cases, the difference between the full and the VBF computation
is minimal and do not exceed few per cent.
Note also that for these distributions, the NNLO non-factorisable
corrections are rather suppressed with small variations over the
kinematic range shown.
They largely inherit the corrections observed at the level of the fiducial cross section.

\begin{figure}
        \setlength{\parskip}{-10pt}
        \begin{subfigure}{0.49\textwidth}
                \subcaption{}
                \vspace{3mm}
                 \includegraphics[width=\textwidth,page=11]{results/VBFHH-plots-paper-extended.pdf}
                \label{plot:mHH}
        \end{subfigure}
        \hfill
        \begin{subfigure}{0.49\textwidth}
                \subcaption{}
                \vspace{3mm}
                 \includegraphics[width=\textwidth,page=13]{results/VBFHH-plots-paper-extended.pdf}
                \label{plot:pTHH}
        \end{subfigure}

        \begin{subfigure}{0.49\textwidth}
                \subcaption{}
                \vspace{3mm}
                 \includegraphics[width=\textwidth,page=17]{results/VBFHH-plots-paper-extended.pdf}
                \label{plot:mjj}
        \end{subfigure}
        \hfill
        \begin{subfigure}{0.49\textwidth}
                \subcaption{}
                \vspace{3mm}
                 \includegraphics[width=\textwidth,page=14]{results/VBFHH-plots-paper-extended.pdf}
                \label{plot:dyjj}
        \end{subfigure}

        \vspace*{-3ex}
        \caption{\label{fig:m}%
          Differential distributions for $\Pp\Pp \to \Pj \Pj \PH \PH$
          at the LHC with centre-of-mass energy of $14\TeV$:
          \subref{plot:mHH}~invariant mass of the two Higgs
          bosons~(top left), %
          \subref{plot:pTHH}~transverse momentum of the two Higgs
          bosons~(top right), \subref{plot:mjj}~invariant mass of the
          two hardest jets~(bottom left), and
          \subref{plot:dyjj}~rapidity difference between the two
          hardest jets~(bottom right).                  The upper panel shows the absolute contributions at NNLO QCD + NLO EW and the LO prediction.
                The lower panel shows the relative corrections. The bands denote the envelope of the QCD scale variation.}
\end{figure}

Finally, in Fig.~\ref{fig:m} the invariant-mass and
transverse-momentum distributions of the di-Higgs system are shown.
For the invariant mass, the EW corrections are positive (above $5\%$)
at $200\GeV$.
In the second and third bin, the maximum of the distribution is
reached.
The corrections become then negative about $-10\%$ at
$2\TeV$.
The QCD corrections on the other hand are rather flat and
largely inherit the overall normalisation.
In the tail of the distribution at about $2\TeV$, the VBF computation
start to diverge from the full one at a level of $5\%$.  The
transverse momentum of the two Higgs bosons displays much larger shape
distortions.  The NLO QCD corrections are maximal at low and high
transverse momentum and are minimal around $200\GeV$.  This originates
from kinematic constraints at low transverse momentum \emph{i.e.}\ in
the first three bins (the cut on the jet transverse momenta being
$25\GeV$).  With higher-order QCD radiation, this constraint is
relaxed allowing a relative increase in the cross section.  At high
transverse momentum, the running of the strong coupling is then taking
place.  Such an effect has already been observed for single Higgs
production \cite{Cacciari:2015jma} and is also visible (even if less
pronounced) in the transverse-momentum distributions of the Higgs
boson in Figs.~\ref{plot:pTH1} and \ref{plot:pTH2}.  The NNLO QCD
corrections slowly increase towards high transverse momentum to $+5\%$
at $600\GeV$.  The EW corrections on the other hand displays a typical
Sudakov behaviour.  The corrections become rather large
($\simeq-25\%$) at $600\GeV$.  But such an energy corresponds to a
rather extreme part of the phase space which is suppressed by more
than three orders of magnitude with respect to the maximum of the
distribution.  Finally, the invariant mass and the rapidity difference
of the two tagging jets are also displayed.  These are typical
observables used by experimental collaborations to enhance the VBF
signal.  Going to high invariant mass, the EW corrections slightly
increase to reach about $-10\%$.  This value is also obtained for
larger rapidity difference of the two tagging jets.  Such a kinematic
(large invariant mass and large rapidity difference) is the typical
VBF kinematic meaning that making the phase space cuts even more
exclusive would increase the EW corrections but not dramatically.
Overall, both at the level of the cross section and for differential
distribution, the findings regarding the EW corrections are very
similar to the ones for single Higgs production
\cite{Ciccolini:2007ec}.

Finally, a few remarks regarding the non-factorisable NNLO QCD
corrections.
For most observable shown here the non-factorisable NNLO QCD
corrections are roughly an order of magnitude smaller than the factorisable NNLO QCD
corrections, with very small shapes differences.
However, one has to be careful as the non-factorisable corrections are
here computed in the eikonal approximation, which is only valid
whenever all transverse momentum scales are much smaller than the
partonic centre-of-mass energy.
In particular this means that the approximation breaks down whenever
the transverse momentum of a jet becomes large.
This is the case in the high-transverse momentum regions of Figs.~\ref{plot:pTj1}-\ref{plot:pTj2}.
For all other observables shown here the eikonal approximation is
expected to be valid as discussed in Ref.~\cite{Dreyer:2020urf}.

\section{Conclusion}

The High-Luminosity phase of the LHC will allow to probe rare
processes, giving insight into fundamental interactions at higher
energies.
One of these rare processes is the production of two Higgs
bosons through VBF.
To that end, SM predictions are critical for the measurements of the
process and associated searches for new physics phenomena.

In the present article we have provided state-of-the-art predictions
within the SM in a realistic experimental set-up at the LHC at
$14\TeV$.
They feature the full LO predictions wit NLO QCD and EW corrections
without relying on any approximation.
These have then been combined with the existing NNLO QCD corrections
to obtain the first predictions at
${\rm NNLO \; QCD \times NLO \; EW}$.
The NLO EW corrections are presented for the first time here as well
as the corresponding state-of-the-art predictions.
Also, some of the differential distributions are shown at NNLO QCD
here for the first time.

We have found that the EW corrections display the typical Sudakov
behaviour in the high-energy limits.
The corrections are of $-6\%$ for the fiducial cross section while
they can typically grow up to $-10\%$ to $-20\%$ in differential
distributions.
In general, the corrections are rather similar to the ones for
single-Higgs productions and do not display very large EW corrections.
The situation is thus rather close to the one of QCD corrections that
largely share similarities between single- and double-Higgs
production.

For the event selection chosen here, we predict a fiducial cross section of
$0.67\fb$.
This amounts to a correction factor of $-14.8\%$ with respect to the
LO predictions.
At the level of the differential distributions, the corrections can be
larger and give rise to shape distortions of the order of
$40\%/50\%$.
We note that for the High Luminosity LHC with an expected integrated
luminosity of $3000\fb^{-1}$ this amounts to more than $2000$ events
in the rather strict fiducial volume used here.

Finally, we have also analysed the differences between the full
computation including $s$-channel contributions and the VBF
approximated one.
This is also the first time that such results are shown in that
respect.
They are in rather good agreement with previous findings for
single-Higgs production.
In line with previous studies for similar processes, we have found
that the VBF approximation becomes unreliable in rather inclusive
set-ups or in extreme regions of the phase space.

The present results provide detailed information regarding higher-order
corrections in double-Higgs production via VBF.
It should thus be used as a guideline by the experimental
collaborations in their quests for the measurement of this process
during the High-Luminosity phase of the LHC.
It can also serve as a reference in the corresponding new-physics
searches for future collider experiments that will help us unravel the
mysteries of fundamental physics.

\section*{Acknowledgements}

We would like to thank Ansgar Denner for useful discussions.
FD is supported by the Science and Technology Facilities Council
(STFC) under grant ST/P000770/1.
AK is supported by the European Research Council (ERC) under the
European Union's Horizon 2020 research and innovation programme (grant
agreement No. 788223, PanScales), and by Linacre College, Oxford.
JNL acknowledges support from the Swiss National Science Foundation (SNF)
under contract BSCGI0-157722.
The research of MP has received funding from the European Research
Council (ERC) under the European Union's Horizon 2020 Research and
Innovation Programme (grant agreement no.~683211).

\bibliographystyle{utphys.bst}
\bibliography{vbfhh}
\end{document}